\documentclass[superscriptaddress,twocolumn,showpacs,pra,longbibliography]{revtex4-1}
\usepackage{bm} 
\usepackage{graphicx} 
\usepackage{amsmath}

\usepackage{amsthm}
\usepackage{amssymb} 
\usepackage{comment}
\usepackage{amsfonts}
\usepackage[normalem]{ulem}
\usepackage{dsfont}
\usepackage{color} 
\usepackage{mathtools}
\usepackage{hyperref}
\usepackage{extarrows}
\usepackage{MnSymbol}
\hypersetup{
    colorlinks=true,       
    linkcolor=cyan,          
    citecolor=magenta,        
    filecolor=magenta,      
    urlcolor=cyan,           
    runcolor=cyan
}
\usepackage{epigraph}

\newcommand{\bra}[1]{\langle #1 |}
\newcommand{\ket}[1]{| #1 \rangle}

\newcommand {\nn}{\nonumber}

\usepackage{soul}

\begin{document}

\title{Accounting for gauge symmetries in CHSH  experiments}
\author{David H. Oaknin}
\affiliation{Rafael Ltd, IL-31021 Haifa, Israel}
\author{Amir Kalev*}
\affiliation{Information Sciences Institute, University of Southern California, Arlington, VA 22203, USA}
\affiliation{Department of Physics and Astronomy, and Center for Quantum Information Science \& Technology, University of Southern California, Los Angeles, California 90089, USA}
\email[Corresponding Author:~]{amirk@isi.edu}
\author{Itay Hen}
\affiliation{Department of Physics and Astronomy, and Center for Quantum Information Science \& Technology, University of Southern California, Los Angeles, California 90089, USA}
\affiliation{Information Sciences Institute, University of Southern California, Marina del Rey, CA 90292, USA}
\date{\today}

\begin{abstract}
\noindent We re-examine the CHSH experiment, which we abstract here as a multi-round game played between two parties with each party reporting a single binary outcome at each round. We explore in particular the role that symmetries, and the spontaneous breaking thereof, play in determining the maximally achievable correlations between the two parties. We show, with the help of an explicit statistical model, that the spontaneous breaking of rotational symmetry allows for stronger correlations than those that can be achieved in its absence. We then demonstrate that spontaneous symmetry breaking may lead to a violation of the renowned CHSH inequality. We believe that the ideas presented in this paper open the door to novel research avenues that have the potential to deepen our understanding of the quantum formalism and the physical reality that it describes.
\end{abstract}

\maketitle

\epigraph{\it `Nothing in physics seems so hopeful
to me as the idea that it is possible for
a theory to have a very high degree of
symmetry which is hidden from us in
ordinary life.'}{{\it Steven Weinberg}~\cite{Weinberg_quote}}

\section{Introduction}
Symmetry is a fundamental concept in modern physics~\cite{Weinberg}. Symmetry transformations encode the covariance of the laws of physics with respect to distinct yet equivalent frames of reference. Modern physical theories are in fact defined by their group of symmetry transformations and by the rules that describe how physical observables transform under these symmetry operations. Moreover, as Noether's theorem reveals, symmetries may further lead to conservation laws. Symmetry considerations may, in certain cases, even determine the general dependence of the statistical correlations between measurable quantities of a physical system~\cite{MUKHANOV1992203}.

Correlations measure the degree to which different features of physical systems fluctuate together. They are responsible for the emergence of complex behavior at all scales, and as such play a key role in virtually all areas of science~\cite{Parisi}. In classical physics, correlations between macroscopic features are understood as emerging from an underlying deterministic microscopic theory, defined on an extended phase-space that includes additional, computationally inaccessible, degrees of freedom. Averaging out these microscopic degrees of freedom leads in turn to a seemingly probabilistic behavior of the remaining degrees of freedom and the correlations thereof. These principles are epitomized in classical statistical physics, which gives rise to a relatively simple description for macroscopic physical systems consisting of a large number of microscopic degrees of freedom~\cite{LL:SP1,LL:SP2,KH:SP1}. 

Bipartite correlations between binary physical quantities appearing in the framework of classical statistical physics are known to be subject to inequality constraints when these are linked together. These constraints stem directly from a result in probabilistic logic known as the Boole-Fr{\'e}chet inequalities, which places bounds on different combinations of probabilities concerning logical propositions or related events~\cite{Hailperin,Frechet}. In physics, Bell's inequalities~\cite{Bell-book,Bell,Fine,CHSH,Klyachko,GHZ,KS} are celebrated examples of this type of constraints, which touch upon the foundations of quantum mechanics and the very nature of physical reality through the renowned Einstein-Podolsky-Rosen (EPR) paradox~\cite{epr,bohm}. 
Bell's inequalities restrict the values of certain linear combinations of correlations between pairs of binary degrees of freedom arranged in a closed loop of paired events. 

A well-known setup in which Bell's inequalities arise is that devised by Clauser, Horne, Shimony and Holt (abbreviated CHSH)~\cite{CHSH}, extending Bohm's version of the EPR paradox~\cite{bohm}. The CHSH setup describes a multi-round game in which two parties, who share at each round of the game a pair of twin particles, can freely and independently control the settings of their respective measurement devices. Upon the detection of their respective particles, the devices produce each a binary outcome. The CHSH version of Bell's inequalities (the CHSH inequality) must hold for any statistical model that aims to describe these games and shares certain physically intuitive features. Nonetheless, the inequality is remarkably violated by the predictions of quantum mechanics for the outcome of the game when it is played with pairs of entangled quantum bits, as well as by the results collected in actual experimental realizations. 

In this study, we revisit the derivation of the CHSH inequality and explore the conditions under which the inequality can be violated in CHSH games~\cite{Frontiers,MPLA}. In particular, we explore the role that symmetries, and the spontaneous breaking thereof, play in determining the bounds on the correlations that can be attained between the measured outcomes of the two parties. We show explicitly that spontaneous symmetry breaking allows for stronger correlations than those attainable in its absence. We accomplish this goal by presenting a statistical model for the CHSH game that possesses the symmetries of the actual CHSH experiment, and in which the response of each detector is local and deterministic. In our proposed model, the rotational symmetry can be, in the general case,  spontaneously broken and as a consequence the correlations between the parties may reproduce those predicted by quantum mechanics and confirmed in experiments~\cite{chsh_experiment_1,chsh_experiment_2,chsh_experiment_3}. Our results can thus be viewed as yet another manifestation of the fundamental role that symmetries, and the breaking thereof, play in physics.

The paper is organized as follows. We begin with reviewing the derivation of the CHSH theorem for the CHSH setup in terms of an inequality in Sec.~\ref{sec:preliminaries}. The aim of this section is to flesh out the specific assumptions that enter into the derivation of the CHSH inequality. In Sec.~\ref{sec:setup} we describe the CHSH experiment and discuss the role that rotational symmetry plays in it. In Sec~\ref{sec:general}, we provide an explicit statistical model for the CHSH experiment. We show that, due to spontaneous symmetry breaking, our model is not necessarily constrained by the prerequisites of the CHSH theorem, and therefore, is not necessarily constrained by its inequality.
In Sec.~\ref{sec:rhoq}, we provide first-principles arguments, rooted in information theory,  as to why Nature might single out `quantum' correlations amongst all other possibilities, including, classical correlations, for the CHSH experiment. We offer a summary and conclusions in Sec.~\ref{sec:con}. 

\section{The CHSH theorem}\label{sec:preliminaries}

The CHSH version of Bell's theorem states that given a probability space $\Omega$ and four binary random variables defined within it, $S_{A_1}, S_{A_2}, S_{B_1}, S_{B_2} \in \left\{-1,+1\right\}$ which have vanishing expectation values, namely, $\langle S_{A_i}\rangle=\langle S_{B_j}\rangle=0$ for $i,j\in\{1,2\}$, the following inequality holds. 
\begin{equation}\label{eq:CHSH_general}
\Big|\langle S_{A_1}S_{B_1}\rangle+\langle S_{A_2}S_{B_1}\rangle+\langle S_{A_2}S_{B_2}\rangle-\langle S_{A_1}S_{B_2}\rangle\Big| \leq 2 \,, 
\end{equation}
where $\langle S_{A_i}S_{B_j}\rangle$ denotes the statistical correlation between the two variables.

The proof of the theorem is straightforward, and is a consequence of the following identity, which must hold for any possible elementary event in $\Omega$:
\begin{align}\label{the_proof}
&\Big| S_{A_1}S_{B_1}+ S_{A_2}S_{B_1}+ S_{A_2}S_{B_2}- S_{A_1}S_{B_2}\Big|\\\nn 
&=\Big| \big(S_{A_1}+ S_{A_2}\big)S_{B_1}+ \big(S_{A_2}- S_{A_1}\big)S_{B_2}\Big|=  2.
\end{align}
The inequality Eq.~(\ref{eq:CHSH_general}) is then obtained by averaging the last identity over the space of possible elementary events. 

\begin{figure}[h]
\includegraphics[width=0.75\columnwidth]{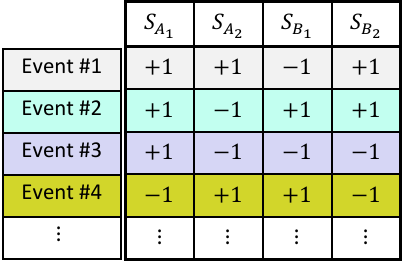}
\caption{\label{fig:ineq_table} {\bf The basic assertion of the CHSH inequality.} The CHSH inequality is based on the assertion that one can assign definite values to the four binary random variables $S_{A_1}, S_{A_2}, S_{B_1}$ and $S_{B_2}$ for any event in $\Omega$. The table illustrates four such events.}
\end{figure}

Importantly, as evident from the proof, the theorem is founded on the prerequisite that all four random binary variables can be attributed a definite binary assignment for any elementary event in $\Omega$. This is illustrated in Fig.~\ref{fig:ineq_table}.   

It is often assumed that the above inequality should trivially hold for any statistical model that aims to describe actual CHSH experiments that are of interest in quantum foundations studies (we describe these in detail in the next section, Sec.~\ref{sec:setup}). However, it is the declared purpose of this paper to point out that the conditions required by the theorem for inequality
Eq.~(\ref{eq:CHSH_general}) to hold, should not necessarily be fulfilled in these CHSH experiments. In particular, we provide a mathematical framework for an explicit family of statistical models that respect all the requirements of the CHSH experiment, but for which the inequality does \emph{not} necessarily hold (this is discussed in Sec.~\ref{sec:general}).

\section{The CHSH experiment}\label{sec:setup}

The CHSH experiment, abstracted here as a game played between two parties, Alice and Bob, requires a source that emits pairs of twin particles. One particle from each pair is sent to Alice and the other to Bob. The location of the source is immaterial for our discussion, however we assume that the parties are located far enough from each other so that the arriving events of the twin particles to their respective parties lay outside each other's causal lightcone.

\begin{figure}[h]
\includegraphics[width=0.75\columnwidth]{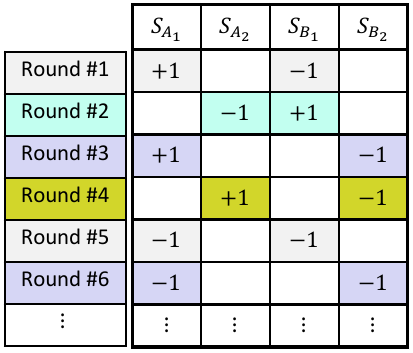}
\caption{\label{fig:exp_table} {\bf A typical output table in a CHSH experiment.} In each round of the game one and only one experimental setting is tested, $({\bm a}_i, {\bm b}_j)$ for $i,j\in\{1,2\}$ and the corresponding binary outputs, $S_{A_i}$ and $S_{B_j}$, are recorded. Unlike the table assumed in order to prove the CHSH theorem, cf. Fig.~\ref{fig:ineq_table} in which all four columns are (or can be) filled out at each round, in the table obtained in a CHSH experiment only two columns are populated at each round.}
\end{figure}

Each party has a `directional' detector, i.e., a device that can be freely oriented along some direction, and that can detect the incoming particles emitted from the source. We will refer to these as detectors $A$ (for Alice) and $B$ (for Bob). To play a CHSH game, Alice and Bob each choose two possible orientations for their respective detectors: ${\bm a}_1$ and ${\bm a}_2$ for Alice and ${\bm b}_1$ and ${\bm b}_2$ for Bob, defined with respect to local lab frames. In each round of the game, the source emits a pair of twin particles and the parties are free to independently set the orientation of their respective detectors along one of their two pre-chosen settings. Upon detection of an incoming particle, each of the detectors produces a binary outcome: $S_A\in\{-1,+1\}$ for detector $A$ and $S_B\in\{-1,+1\}$ for detector $B$. We emphasize that in each round of the game one, and only one, of the four possible joint settings of the pair of detectors $({\bm a}_i,{\bm b}_j)$ with $i,j\in\{1,2\}$ is (and can be) tested, and therefore only two binary $S_{A_i}, S_{B_j}$ outcomes are obtained, one per party (see Fig.~\ref{fig:exp_table}).

Upon a (joint) detection event, the parties record the bits of information output by their respective detectors, along with the detector settings. At the end of the game, after many rounds, the two parties use the experimental data to evaluate the bipartite correlations for each of the four joint settings $({\bm a}_1,{\bm b}_1), ({\bm a}_1,{\bm b}_2), ({\bm a}_2,{\bm b}_1)$ and $({\bm a}_2,{\bm b}_2)$. For any one of the four joint settings, the expected (empirical) bipartite correlation $\langle S_{A_i}S_{B_j}\rangle$, for $i,j=1,2$, is defined as the difference between the probability (fraction of events) that the two parties report the same outcome and the probability (fraction of events) that they report opposite outcomes. As a last step, the score of the game is defined by 
\begin{equation}\label{eq:score}
\text{Score}=\Big|\langle S_{A_1}S_{B_1}\rangle+\langle S_{A_2}S_{B_1}\rangle+\langle S_{A_2}S_{B_2}\rangle-\langle S_{A_1}S_{B_2}\rangle\Big|. 
\end{equation}

Importantly, we note that in CHSH experiments, the particle-emitting source does not have any preferred direction: The expectation value of each detector's outcomes is zero, irrespective of its orientation, while the correlation between the outcomes of the two devices is a function only of their relative orientation. 

It may seem trivial at first to expect that the statistical correlations between the outcomes of the two detectors at the four joint settings should satisfy the conditions of the CHSH theorem presented above. After all, even though in each round of the game only two binary outcomes are produced, the event space of the experiment seems to be defined through four binary random variables $S_{A_1}, S_{A_2}, S_{B_1}$ and $S_{B_2}$ and four joint settings $({\bm a}_1,{\bm b}_1), ({\bm a}_1,{\bm b}_2), ({\bm a}_2,{\bm b}_1)$ and $({\bm a}_2,{\bm b}_2)$ (regardless of whether they have been measured or not), which appear to be the only conditions required for the CHSH inequality, Eq.~\eqref{eq:CHSH_general}, to hold.

Contrary to this expectation, Quantum Mechanics (QM) predicts~\cite{Peres:book} -- and carefully performed experiments confirm~\cite{chsh_experiment_1,chsh_experiment_2,chsh_experiment_3} -- a violation of the CHSH inequality in CHSH games played with pairs of qubits prepared in a maximally entangled state. Specifically, QM predicts that the average detection outcome of each detector is zero, irrespective of its orientation, while the statistical correlation between the outcomes of the two detectors is given by 
\begin{equation}\label{qmprediction}
 \langle S_A S_B\rangle = -\cos({\theta_{AB}}),   
\end{equation}
where   $\theta_{AB}$ is the relative angle between the detectors with respect to a setting where their experimental results are fully anticorrelated.  Even though this expression holds for any relative orientation of the detectors defined over the 3D unit spheres, we shall assume here, for the sake of simplicity and without any loss of generality, that the relative orientation of the two detectors is constrained to a calibrated plane, so that it is completely specified by a single $2\pi$-periodic angular degree of freedom, i.e., a point on the unit circle.

For reasons that will become clearer below, we notice that, more generally, we can express the relative orientation between the detectors as $\theta_{AB}=\Phi+\vartheta_{AB}$, where $\vartheta_{AB}$ denotes the relative angle between the detectors with respect to a setting where the correlation between their outcomes is $-\cos({\Phi})$ (in Appendix~\ref{app:quant} we provide a detailed discussion of the decomposition of the angle in the context of Quantum Mechanics).  This trivial observation places emphasis on the fact that consecutive co-planar rotations enter the correlation in an additive manner, so that any relative setting of the detectors can be taken as a reference to define a subsequent relative rotation, on the same footing as the setting in which their outcomes are fully anti-correlated. Hereafter, for simplicity we refer to $\theta_{AB}$ as the relative angle between the detectors, notwithstanding the above nuance. 

Accordingly, for a setup of four joint settings of the detectors at relative angles $-\theta_{A_2 B_1} = \theta_{A_2 B_2} = \theta_{A_1 B_1} = \theta_{A_1 B_2} - \frac{\pi}{2} = \frac{\pi}{4}$, the expected correlations are $\langle S_{A_2} S_{B_1}\rangle = \langle S_{A_2} S_{B_2}\rangle = \langle S_{A_1} S_{B_1}\rangle = -\langle S_{A_1} S_{B_2}\rangle = 1/\sqrt{2}$, so the 
expected score (\ref{eq:score})
\begin{equation}\label{eq:scoreQM}
\text{Score} =  2 \sqrt{2},    
\end{equation}
violates the CHSH inequality obtained in Sec.~\ref{sec:preliminaries}. 

This violation, which has been widely understood for decades as a unique feature of quantum theory, is of course not in contradiction with the mathematical statement of the CHSH theorem. It simply implies that the underlying assumptions of the theorem are not met in QM, nor they are met in the experiment. In particular, the derivation of the inequality Eq.~\eqref{eq:CHSH_general} demands  assigning four binary outcomes ($S_{A_1},S_{A_2}, S_{B_1}$ and $S_{B_2}$, regardless of whether they have been measured or not) to each elementary event in the space of events that may take place in each round of the experiment. This demand requires, in turn, identifying each detector's setting, ${\bm a}_1,{\bm a}_2,{\bm b}_1$, and ${\bm b}_2$, as a physically meaningful label. Nonetheless, as we shall discuss in detail throughout this paper, neither within the framework of QM nor in actual CHSH experiments it is possible -- due to gauge symmetry considerations -- to associate ${\bm a}_1,{\bm a}_2,{\bm b}_1$, and ${\bm b}_2$ with unique physical degrees of freedom, as these are merely (arbitrary) labels defined with respect to Alice and Bob's labs, respectively. 

Before introducing our statistical model, in which gauge symmetries play a crucial role as was just mentioned, for context, we find it useful to first discuss the role that gauge symmetries play in our current understanding of the microscopic world. In high-energy particle physics and condensed matter physics, gauge symmetries correspond to locally defined symmetries that encode the interactions allowed in the system. In general, gauge symmetries are associated with spurious degrees of freedom that may appear in the theoretical formalism but do not correspond to actual physical degrees of freedom. These were already identified in the early days of the development of quantum mechanics~\cite{Dirac_book}. A particularly striking example of the role played by gauge symmetries can be found in the theoretical description of the swimming of microscopic bodies at low Reynolds number~\cite{Wilczek}. 

In the present work, we point out that gauge symmetries may play also a crucial role in understanding the physics of CHSH experiments, in which a violation of the CHSH inequality is obtained.  To do so, in the next section we  consider statistical models that share with the quantum formalism the key property that  joint detectors settings characterized by the same relative orientation are physically indistinguishable and are related to each other by gauge rotations. We shall show that, similarly to the quantum formalism, our statistical models are not necessarily constrained by the prerequisite of the CHSH theorem, and therefore, are not limited by its inequality.
 
\section{A statistical model for the CHSH game}\label{sec:general}

We next present an explicit statistical model for the CHSH experiment that as we show can reproduce the predictions of Quantum Mechanics and the data collected in actual experiments. As we will demonstrate, our proposed model possesses all the required symmetries, i.e., the  expectation values of the outcomes of each detector are zero, irrespective of the detector orientation, while the correlation between the outcomes of the two devices is a function only of their relative orientation.  On the other hand, as we shall see, the model does not meet the requirements underlying the derivation of the CHSH theorem, and as such it is not necessarily constrained by the CHSH inequality, Eq.~(\ref{eq:CHSH_general}). Our model builds on the observation made above, that the relative orientation between the two detectors, $A$ and $B$, completely defines the setting of the experiment, while the orientations of each one of the detectors with respect to local lab frames do not (and must not) play a role in the physical description of the experiment. 

To capture these elements in our statistical model, we consider a source that produces randomly oriented pairs of twin particles, sent to the two parties, one particle to each participant. In the absence of an absolute frame of reference, it is  meaningless to assign a distinct physical degree of freedom to the orientation of these objects without relating them to the orientation of the corresponding detectors, as much as it is meaningless to label an absolute orientation to a detector without relating it to the orientation of the other detector.   Hence, in our model, we label the incoming particles only through their  relative orientation to the two detectors, $A$ and $B$, with the help of corresponding random variables defined over the unit circle,  $\Delta_A$ and $\Delta_B$. \footnote{Formally, a random variable is a \emph{function} from the sample space of a probability space to a measurable space. In our model, the sample space is (and can only be) described with respect to either one of the observers, however not in absolute terms. Thus, formally the random variable is the identity mapping $\Delta_N(t)=t$ (where $t\in[-\pi,\pi)$) for either $N=A$ or $N=B$. Since  the random variable is the identity mapping, in order to avoid unnecessary heavy notation, in what follows we will not be carrying $t$ in our equations and treat $\Delta_N$ (for $N=A$ or $N=B$) as a variable.} By doing so, we refrain from using `absolute' labels and relations to arbitrarily chosen lab frames of reference. By convenience, we define the zero of the random variables $\Delta_A$ and $\Delta_B$ along the orientation of the detectors $A$ and $B$, respectively.

We further allow for the probability density of these random variables $\Delta_A$ and $\Delta_B$ to be non-uniform over their domain of definition $\Delta_A, \Delta_B \in [-\pi,\pi)$. The key insight, however, is that in order to meet the symmetry requirements described above, we must design our statistical model such that both random variables have the same non-uniform probability density function, $\rho(\Delta_A)$ and  $\rho(\Delta_B)$, independently of the relative orientation between the two detectors. This ensures that  our model satisfies all the symmetry constraints of the CHSH experiment while not imposing any physically-unjustified constraints, such as those derived from a choice of an external reference frame.

In addition, following the same symmetry requirements of the actual experiments, we restrict ourselves to cases where the probability density is parity symmetric, that is, \hbox{$\rho(-\Delta)=\rho(\Delta)=\rho(\pi-\Delta)$}, $\forall \Delta \in [0,\pi]$. We further demand that the probability density is not identically zero on any subinterval of $[-\pi,\pi)$. These demands guarantee the existence of a strictly monotonically increasing (i.e., bijective) mapping $\Gamma^{-1}:[-\pi,\pi)\rightarrow[-\pi,\pi)$ such that
\begin{align}\label{eq:Gamma-1}
\Gamma^{-1}(\Delta) = 
\begin{cases}
    2\pi \int_0^{\Delta} d\xi \ \rho(\xi),&\text{if~} \Delta \in [0, \pi)\\
    -2\pi \int_{\Delta}^0 d\xi \ \rho(\xi), &\text{if~} \Delta \in [-\pi, 0)
\end{cases}.
\end{align}
Since by definition, $\Gamma^{-1}$ is bijective and anti-symmetric, $\Gamma^{-1}(-\Delta)=-\Gamma^{-1}(\Delta)$, its inverse $\Gamma(\cdot)$ is bijective and anti-symmetric as well. Therefore, in particular, $\Gamma(0)=0$ and $\Gamma(\pm \pi)=\pm \pi$. Moreover,  one can show that $\Gamma^{-1}$ preserves orthogonality, $\Gamma^{-1}\left(\pm\frac{\pi}{2}\right)=\pm\frac{\pi}{2}$ (and, hence, also $\Gamma(\cdot)$ does). Most importantly for our discussion, if $\Delta$ is a variable randomly distributed in $[-\pi,\pi)$ with a probability density $\rho(\Delta)$ then the random variable $ \Gamma^{-1}(\Delta)$ is distributed over the same range with a uniform probability density. Similarly, if $\Delta$ is a uniformly distributed random variable in $[-\pi,\pi)$, then $\Gamma(\Delta)$ is distributed according to a probability density $\rho(\Delta)$.

Using $\Gamma(\cdot)$ and its inverse we can now write the relation
\begin{equation}\label{eq:trinity}
\Delta_{A}=\Gamma\big(\Gamma^{-1}(\theta)-\Gamma^{-1}\big(\Delta_{B}\big)\big),
\end{equation}
where the addition $\Gamma^{-1}(\theta)-\Gamma^{-1}\big(\Delta_{B}\big)$ should be understood as  defined modulo $2\pi$ over the interval $[-\pi,\pi)$. 

The angle $\theta\in[-\pi,\pi)$ in Eq.~\eqref{eq:trinity} has exactly the same interpretation as the angle denoted as $\theta_{AB}$ in the previous section: it describes the relative orientation of the detectors with respect to a setting in which their outcomes are fully anticorrelated, and can be generally defined as $\theta=\Phi+\vartheta$, where $\vartheta$ is a relative orientation of the detectors with respect to a reference setting in which the correlation between the outcomes of the  detectors is characterized by an angle $\Phi$. The constant $\Gamma^{-1}(\theta)$ is determined by the requirement that when  $\Delta_{B}=0$ we must have $\Delta_A=\theta$, and vice versa. 

As a consequence, in our model the algebra associated with consecutive relative rotations of the two detectors is linear, despite the non-linear transformations of Eq.~\eqref{eq:trinity}. This feature (i.e., the additivity of co-planar rotations of the detectors' relative orientation) is derived in our model from the fundamental observation that any relative setting of the detectors can be taken as a reference to define a subsequent rotation, on the same footing as the setting in which their outcomes are fully anticorrelated. Equation~(\ref{eq:trinity}) can be readily generalized to take into account non-coplanar rotations as well (see Appendix \ref{app:extSph}).

Furthermore, Eq.~\eqref{eq:trinity} ensures that if $\Delta_{B}$ is distributed according to the probability density $\rho$ then so does $\Delta_{A}$, and vice versa. This can be readily deduced  by noting that given that $\Delta_{B}$ is distributed according to $\rho$ then $\Gamma^{-1}\big(\Delta_{B}\big)$ is uniformly distributed. Adding the  constant $\Gamma^{-1}(\theta)$ (modulo $2\pi$) to  $\Gamma^{-1}\big(\Delta_{B}\big)$ keeps the distribution uniform in $[-\pi,\pi)$. Therefore, by applying $\Gamma(\cdot)$ on the right-hand-side of Eq.~\eqref{eq:trinity} to a uniformly distributed random variable, we obtain that $\Delta_{A}$ is distributed according to $\rho$, as required. This specifically guarantees that the probability measure $\rho(\Delta) d\Delta$ is preserved under the transformation defined by Eq.~\eqref{eq:trinity}, i.e., $\rho(\Delta_A) d\Delta_A =  \rho(\Delta_b) d\Delta_B$, and hence  Eq.~\eqref{eq:trinity} assures measurement independence, that is, that the measurement setting made by an observer is (statistically) independent of the properties of the system being measured~\cite{hall11}. 

Equation~\eqref{eq:trinity} is the main technical ingredient of the proposed model. It states that the symmetries of the CHSH experiment do {\it not} necessarily imply that $\theta$, $\Delta_{A}$ and $\Delta_{B}$ are related via a linear transformation law (as one may expect naively), but rather demand that the source-generated twin particles appear with the same distribution $\rho(\cdot)$ with respect to detector $A$ {\it and} with respect to detector $B$, irrespective of their relative orientation $\theta$. \footnote{This equation can be heuristically understood as a {\it principle of relativity}, that to an extent facsimiles Einstein's observation regarding the invariance of the physical laws with respect to all observers, either inertial or not. In the context of our model, we notice that all settings of the detectors, either parallel (fully anti-correlated) or not, may serve on equal footing as a reference setting to describe the source of pairs of twin particles as well as subsequent relative rotations of the detectors.}

Upon detection of the source-generated pair, each detector produces a binary outcome according to the rule 
\begin{align}\label{eq:responseN_standard}
S_N(\Delta_{N}) = \begin{cases}
-1&\text{if~}\Delta_{N} \in [-\pi, 0),\\
    +1&\text{if~}\Delta_{N} \in [0, \pi),    \end{cases}
\end{align}
with $N\in\{A,B\}$. This response is deterministic and local, that is, the output of each detector is determined uniquely by the relative orientation $\Delta_N$ of the incoming particle, and it is independent of the orientation of the detector of the other party.

Since the probability density $\rho(\Delta)$ is parity-symmetric it immediately follows that the expected average of the binary outcomes produced by the detectors at each location is zero, independently of the relative orientation between the two, i.e., 
\begin{align}
\langle S_A \rangle = \langle S_B \rangle = \int d{\Delta_N}\ \rho(\Delta_N) \  S_N(\Delta_N) \ = 0,
\end{align}
consistently with the symmetry conditions discussed above.

We are now in a position to calculate the correlation between Alice and Bob's outcomes in this model. It is given by
\begin{align}\label{eq:corr_general}
&\langle S_{A} S_{B}\rangle\\\nn&= \int d{\Delta_{A}}\ \rho(\Delta_{A}) \ S_A(\Delta_{A}) \ S_B\big(\Gamma(\Gamma^{-1}(\theta)-\Gamma^{-1}(\Delta_{A}))\big)\\\nn
&= \int d{\Delta_{B}}\ \rho(\Delta_{B}) \  S_A\big(\Gamma(\Gamma^{-1}(\theta)-\Gamma^{-1}(\Delta_{B}))\big) \ S_B(\Delta_{B}) \,,
\end{align}
where  we made use of the relation given by Eq.~\eqref{eq:trinity}. 

An explicit expression for the correlation function is obtained by noting that the response of detector $A$ changes sign when $\Delta_{A}=0$ and $\Delta_{A}=\pi$ while the response of detector $B$ changes sign when $\Delta_{A}=\theta$ and $\Delta_{A}=\theta\mp\pi$ (modulo $2\pi$),
\begin{align}\label{eq:cut_standard}
\Delta_{A}=\theta \ \ &\Leftrightarrow \ \  \Delta_{B}=0,\\\nn
\Delta_{A}=\theta\mp\pi \ \ &\Leftrightarrow \ \  \Delta_{B}=\pm \pi.
\end{align}
Since the expected correlation between the outputs of the detectors is given by the difference between the probability that they report the same outcome and the probability they report an opposite outcome [and $\Delta_{A}$ and $\Delta_{B}$ are  distributed according to $\rho(\cdot)$], we can  write the correlation between Alice and Bob, Eq.~\eqref{eq:corr_general}, as:
\begin{align}\label{eq:corr_0_pi}
&\langle S_{A} S_{B}\rangle= \int_{-\pi}^{\theta-\pi} d\xi\ \rho(\xi) -  \int_{\theta-\pi}^0 d\xi\ \rho(\xi)\\\nn
&+ \int_0^{\theta} d\xi\ \rho(\xi) - \int_{\theta}^{\pi}d\xi\ \rho(\xi)=-1 + \frac2{\pi}\Gamma^{-1}(\theta), 
\end{align}
for $\theta\in[0,\pi]$, and 
\begin{align}\label{eq:corr_-pi_0}
&\langle S_{A} S_{B}\rangle=-\int_{-\pi}^{\theta} d\xi\ \rho(\xi) +  \int_{\theta}^0 d\xi\ \rho(\xi) \\\nn
&- \int_0^{\theta+\pi} d\xi\ \rho(\xi) + \int_{\theta+\pi}^{\pi} d\xi\ \rho(\xi)=-1 - \frac2{\pi}\Gamma^{-1}(\theta), 
\end{align}
for $\theta\in[-\pi,0]$. Thus, we find that
\begin{align}\label{eq:corr_general1}
E^{\rm (\rho)}_{AB}(\theta) \coloneq \langle S_{A} S_{B}\rangle &=-1 + \frac2{\pi}|\Gamma^{-1}(\theta)|{=}-1+4\big|\int_0^{\theta}d\xi\ \rho(\xi)\big| 
\end{align}
for $\theta\in[-\pi,\pi)$. As desired, the bipartite correlation is an even function of the relative angle between the detectors and it satisfies
\begin{align}\label{calibration_curve}
E^{\rm (\rho)}_{AB}(0)=-1,\;\;
E^{\rm (\rho)}_{AB}\Bigl({\pm \frac{\pi}{2}}\Bigr)=0,\;\;
E^{\rm (\rho)}_{AB}(\pm \pi)= 1.
\end{align}
By properly choosing the probability density $\rho$, our model is capable of reproducing any bipartite correlation function as long as it fulfills the symmetry constraint
$E^{\rm (\rho)}_{AB}(\theta)=E^{\rm (\rho)}_{AB}(-\theta)=-E^{\rm (\rho)}_{AB}(\pi-\theta)$. 

For example, by considering $\rho(\Delta)=1/2\pi\coloneqq \rho_{\rm cl}(\Delta)$, that is, a uniform distribution over $[-\pi,\pi)$, we obtain the bipartite correlation 
\begin{align}\label{eq:corr_standard}
E^{(\rho_{\rm cl})}_{AB}(\theta) = -1 + \frac2{\pi}|\theta| 
\end{align}
for $\theta\in[-\pi,\pi)$, which corresponds to the strongest correlation function allowed by the  CHSH inequality, Eq.~\eqref{eq:CHSH_general}.  

Alternatively, by considering the probability density function
\begin{align}\label{quantum_rho}
\rho(\Delta) = \frac{1}{4}\left|\sin\left(\Delta\right)\right|\coloneqq\rho_{\rm q}(\Delta),
\end{align}
we arrive at bipartite correlations of the form
\begin{align}\label{eq:corr_quantum}
E^{\rm (\rho_q)}_{AB}(\theta)=-\cos(\theta) \,,
\end{align}
which reproduces the correlation dependence predicted by quantum mechanics for the CHSH experiment with pairs of maximally entangled qubits, Eq.~\eqref{qmprediction}.  We note by passing that while $\rho_q$ corresponds to maximally entangled qubit pair (in terms of the established correlations), there does not exist a one to one correspondence between quantum states and permissible probability densities $\rho$'s in our model.

A violation of the CHSH inequality is possible because the statistical model, by construction, does not fulfill, in general, the requirements for the CHSH theorem.
As discussed in Sec.~\ref{sec:preliminaries}, the CHSH inequality is derived under the premise that four random binary variables $S_{A_1}, S_{A_2}, S_{B_1}$ and $S_{B_2}$ take on well-defined values at each round of the game, irrespective of how many of them are actually logged. This premise is not  fulfilled in our statistical model, as we now explicitly show. 

To further illustrate the key properties of our proposed model,  let us consider the following thought experiment carried out at each round of the game. Let detectors $A$ and $B$ be at a relative angle $\theta_{A_1B_1}$. In this setting, let  $\Delta_{A_1}$ and $\Delta_{B_1}$ be, respectively, the relative angle between the pair produced at one round of the game and each one of the detectors. According to Eq.~\eqref{eq:trinity}, the two are related by $\Delta_{B_1}=\Gamma(\Gamma^{-1}(\theta_{A_1B_1})-\Gamma^{-1}(\Delta_{A_1}))$. Next, suppose that Alice rotates her detector relative to Bob's such that the relative orientation between detector $A$ and detector $B$ is now $\theta_{A_2B_1}$. Since the angle $\Delta_{B_1}$ does not change by Alice's rotation, we obtain that the relative angle between the pair produced at the considered instance and the (new) orientation of detector $A$ is given by 
\begin{align}
\Delta_{A_2}&=\Gamma(\Gamma^{-1}(\theta_{A_2B_1})-\Gamma^{-1}(\Delta_{B_1}))\\\nn
&=\Gamma(\Gamma^{-1}(\theta_{A_2B_1})-\Gamma^{-1}(\theta_{A_1B_1})+\Gamma^{-1}(\Delta_{A_1})).
\end{align}
Next, let Bob change the orientation of his detector so that it is oriented at a relative angle $\theta_{A_2B_2}$ to Alice's.  Again, since the angle $\Delta_{A_2}$ is not altered by this rotation, the relative angle between the pair and detector $B$ is now given by $\Delta_{B_2}=\Gamma(\Gamma^{-1}(\theta_{A_2B_2})-\Gamma^{-1}(\theta_{A_2B_1})+\Gamma^{-1}(\theta_{A_1B_1})-\Gamma^{-1}(\Delta_{A_1}))$.  Last, suppose that Alice changes the orientation of her detector back to the original relative setting with respect to Bob's, $\theta_{A_1B_2}$. Upon denoting by $\Delta_{A_1}^{\!\circlearrowleft}$ the  relative angle between the orientation of detector $A$ and the considered pair, we arrive at
\begin{align}\label{eq:40}
\Delta_{A_1}^{\!\circlearrowleft}&=\Gamma\big(\Gamma^{-1}(\theta_{A_1B_2})-\Gamma^{-1}(\theta_{A_2B_2})\\\nn&+\Gamma^{-1}(\theta_{A_2B_1})-\Gamma^{-1}(\theta_{A_1B_1})+\Gamma^{-1}(\Delta_{A_1})\big)\\\nn
    &=\Gamma\big(\Gamma^{-1}(\delta)+\Gamma^{-1}(\Delta_{A_1})\big),
\end{align}
with $\delta=\Gamma\big(\Gamma^{-1}(\theta_{A_1B_2})-\Gamma^{-1}(\theta_{A_2B_2})+\Gamma^{-1}(\theta_{A_2B_1})-\Gamma^{-1}(\theta_{A_1B_1})\big)$. 
Due to the non-linearity of $\Gamma^{-1}$, in general $\delta\neq0$ (unless $\rho=1/2\pi$), and hence, $\Delta_{A_1}^{\!\circlearrowleft}\neq\Delta_{A_1}$. 
This thought experiment showcases one crucial property of the proposed model: While $\Delta_{A_i}$ and $\Delta_{B_j}$ --- the relative orientations between any source-generated pair and the two detectors $A$ and $B$ positioned at a relative experimental angle $\theta_{A_iB_j}$, with $i,j\in\{1,2\}$ --- are well defined quantities related via Eq.~\eqref{eq:trinity}, it is in principle impossible within our model -- and indeed not required to describe the physics of actual CHSH experiments -- to consistently define concurrently $\Delta_{A_1},\Delta_{A_2},\Delta_{B_1}$ and $\Delta_{B_2}$, due to a {\it holonomy} rooted in the acquired geometric phase $\delta$. Therefore, while two binary variables $(S_{A}, S_{B})$ can be assigned a definite value for any relative orientation of the detectors, it is impossible in our (local, deterministic) model to assign definite values to four binary variables $S_{A_1}, S_{A_2}, S_{B_1}$ and $S_{B_2}$ at each instance of the game (unless $\rho=1/2\pi)$. 

The appearance of the nontrivial geometric phase $\delta$ in Eq.~(\ref{eq:40}) acquired through a closed sequence of coordinate transformations, can be traced back to the symmetry conditions  that preserve  $\rho(\Delta)$ over the unit circle with respect to both detectors $A$ and $B$ irrespective of their relative orientation. We emphasize that this phase is allowed because, as noted above, a joint rotation of the two detectors is associated with an unphysical gauge degree of freedom, and in any instance of the game at most two binary outcomes can be registered at once. Moreover, the existence of this phase is not in contradiction with any feature of actual CHSH experiments, since it preserves all symmetry requirements. Interestingly, this geometric phase can only be indirectly inferred through the emergence of nontrivial correlations between the binary outcomes of the two parties. 
 
The characteristic feature of our model in which  individual configurations transform into each other, namely $\Delta_A \longrightarrow \Delta_A^{\!\circlearrowleft}$, over a closed sequence of relative rotations of the detectors, while their respective probability density functions remain intact, namely $\rho(\Delta_A)$ and $\rho(\Delta_A^{\!\circlearrowleft})$ respectively, is reminiscent of the notion of spontaneous breaking of a symmetry in quantum field theories. We therefore call it here by analogy {\it spontaneous breaking of the gauge rotational symmetry}. 

We stress, in particular, that the gauge symmetry is statistically restored on average over a sequence of many repetitions so that correlations do only depend on the relative orientation of the detectors and, therefore, it is consistent with Elitzur's theorem~\cite{Elitzur}, which states that expectations values in quantum field theories must not depend on any gauge degree of freedom.

\section{Nature's choice of quantum correlations}\label{sec:rhoq}

Other than the generic symmetry constraints listed in Eq.~\eqref{calibration_curve}, the probability density function $\rho(\Delta)$ is a free input in our model. This freedom calls forth the question: Why does Nature favor the ``quantum correlations'', Eq.~\eqref{eq:corr_quantum}, over all other possibilities in the space of all allowed correlations, which include among them also ``classical'' correlations, Eq.~\eqref{eq:corr_standard}? Is there a physical principle that singles out the choice $\rho_{\rm q}(\Delta)$, Eq.~\eqref{quantum_rho}?

While a detailed consideration of this question is beyond the scope of this study, we provide here a few key insights that may shed some light on the answer. Suppose that Alice and Bob would like to estimate the relative angle $\theta$ between the orientation of their detectors with the help of the source of particles. They do so by estimating the correlation between their experimental results, and use this correlation as an (unbiased) estimator for $\theta$. The result in each round of the experiment is a random variable that can take on one of two values: $+1$ if their detectors produce the same binary outcome and $-1$ if they produce opposite outcomes. The probability for the two possible outcomes of this Bernoulli random variable contains  information about $\theta$, as shown in Eqs.~\eqref{eq:corr_0_pi} and~\eqref{eq:corr_-pi_0}:
\begin{equation}
p_{+}(\theta) = 2\int_0^{\theta} d\Delta~\rho(\Delta),\,~\text{and}~p_{-}(\theta)=1-p_{+}(\theta),
\end{equation}
where we have assumed for the sake of concreteness that $\theta$ lays in the interval $[0, \pi)$ (since, by symmetry considerations, we cannot distinguish between $\theta$ and $-\theta$).

The Fisher information, ${\cal I}$, that this random variable carries about $\theta$ is by definition~\cite{Luo} 
\begin{equation}
{\cal I}(\theta) \equiv \sum_{x=\pm} p_x(\theta) \left(\frac{\partial \ \mbox{ln}[p_x(\theta)]}{\partial \theta}\right)^2,     
\end{equation}
which, in our model, equals to
\begin{equation}
\label{Fisher}
{\cal I}(\theta)= \frac{4 \rho^2(\theta)}{\left(2 \int_0^{\theta} d\xi \rho(\xi) \right) \left(1 - 2 \int_0^{\theta} d\xi \rho(\xi) \right) }\coloneqq{\cal I}_{\rho}(\theta).
\end{equation} 
Our main observation is that the Fisher information ${\cal I}_{\rho}(\theta)$ is constant, i.e., independent of $\theta$, for Nature's choice:
\begin{equation}
\label{Fisher_q}
{\cal I}_{\rho_q}(\theta) = 1.
\end{equation}
The implication of this result can be understood from the Cram{\'e}r-Rao bound~\cite{cramer,rao} which sets the fundamental relation between the Fisher information and the precision with which the parameter $\theta$ can be estimated from a sample of experimental results. For Nature's choice this relation yields 
\begin{equation}
\sigma^2_{\rho_{\rm q}}(\theta) \geq \frac{1}{{\cal I}_{\rho_q}(\theta)} = 1 \,,
\end{equation}
thus implying that Nature's choice is the only one that produces `informational' rotational symmetry: an invariance of this precision for all possible experimental settings $\theta$. Interestingly, according to our model, in order to achieve rotational symmetry in this sense, the gauge symmetry must be spontaneously broken by the non-uniform $\rho_{\rm q}(\Delta)$. 

These conclusions may be also inferred directly from a remarkable result by Luo~\cite{Luo}, where it was shown  that the following three statements about the Fisher information of a  Bernoulli random variable are equivalent: (1) The average Fisher information is minimum, assuming there is not any prior preference for any value of the estimated parameter; (2) The Fisher information is constant with respect to the estimated parameter, i.e., invariant in $\theta$ for the case of our interest; (3) The probability for the two possible outcomes of the Bernoulli variable are given by $p_{+}(\theta)=\sin^2(\theta/2)$ or equivalently $\rho(\Delta)=\rho_{\rm q}(\Delta)$ of Eq.~\eqref{quantum_rho}. The model we described in this paper offers a space of allowed probability densities ($\rho(\Delta)$) among which these information theoretic considerations single out Nature's choice $\rho_{\rm q}(\Delta)$. 

Following the results by Luo~\cite{Luo}, these observations may also be related to the {\it principle of minimum Fisher information} proposed by Frieden {\it et al.}~\cite{Frieden1, Frieden2}. The proposed principle states that under some regular conditions and constraints, Nature will adopt a probability density over the available phase space having minimum average Fisher information. The intuition behind this principle is inherently related to information and estimation theory by assuming that Nature's choices correspond to those that grants us the minimal average information regarding unknown parameters of physical systems~\cite{Luo}.

\section{Summary and conclusions}\label{sec:con}

In this work, we studied the bipartite correlations that emerge in statistical models that aim to describe two-party CHSH games. We showed that by relaxing the  implicitly assumed existence of an absolute reference frame relative to which angles are defined, classical model that respects the principles of locality, determinism, and free will can give rise to correlations that violate the celebrated CHSH inequalities. 
By doing so, we offered a physically intuitive resolution to the EPR paradox and the correlations associated with quantum entanglement. 

We note that our model does not contradict the CHSH theorem~\cite{CHSH} --  our setup simply does not satisfy the assumption of a preferred reference frame that is underlying the theorem. Our model is founded upon the simple observation that the bipartite correlations in  the CHSH experiment depend only on the relative orientation between the  detectors of the experiment, while the angle associated with a joint rotation of the two detectors does not (and must not) play a role in the prediction of physical observations. Thus, this degree of freedom should be treated as a spurious gauge degree of freedom in any physical model for the experiment. By treating it as such in our model, we are given the freedom to consider a situation where rotational symmetry is spontaneously broken (for each one of the parties) in each realization of the experiment. This in turn led to the appearance of a nonzero geometric phase through cyclically concatenated coordinate transformations, which in turn led to a violation of the requirements needed for the CHSH theorem to hold. 
It is important to note that on average, over the course of many rounds, the rotational symmetry associated with joint rotations of the two detectors is restored, as it should, and the appearance of a nonzero geometric phase becomes evident only through the (nontrivial) correlations measured in the experiment. 

We believe that the theoretical ideas presented in this work open the door to novel avenues of research that have the potential to deepen our understanding of the quantum formalism and the physical reality that it describes. In particular, the framework presented here offers a physically intuitive explanation to the renowned Tsirelson's bound~\cite{Tsirelson} imposed by the quantum formalism on CHSH's correlations, which seems to be related to symmetry considerations and the so-called principle of minimum Fisher information~\cite{Frieden1,Frieden2,Luo}.

Last, we point out a conceptual resemblance between the framework developed here and the fundamental principles underlying General Relativity. In our framework each detector has  its own associated local set of coordinates, with respect to which it describes the orientation of its incoming particles, while the sets of coordinates associated to the two detectors involved in the experiment are related to each other by a coordinates transformation set by their relative orientation $\theta$. This physical degree of freedom encapsulates within it the setting of the detectors as a relative rotation by an angle $\vartheta$ with respect to a calibration setting characterized by a relative phase $\Phi$~ ($\theta=\vartheta+\Phi$). Importantly, we noted that the correlation between the outcomes of the two detectors in the final setting does not depend individually on the two angles $\vartheta$ and $\Phi$ that describe the calibration process, but only on their sum $\theta$. In this sense, all different choices for the calibration stage are gauge equivalent. A similar situation occurs in the calculation of cosmological redshifts in General Relativity. In these calculations, the observable redshift can be written as consisting of gravitational and Doppler contributions. These two contributions, nonetheless, are not uniquely defined and their respective contributions are set only through a gauge calibration process~\cite{Chodorowski2011}. We plan to explore this conceptual parallel in more depth in  future work.
\\

\bibliography{refs}

\appendix 
\section{Quantum description of two-qubit maximally entangled states }\label{app:quant}
For the sake of completeness, in this appendix we briefly summarize the QM description of a two-qubit maximally entangled state. This is done in order to highlight the features of the entangled state that were required to replicate in the building of the statistical model described in Sec.~\ref{sec:general}. 

To begin with, let us consider the singlet state, conventionally written in the form 
\begin{equation}\label{wavefunction}
|\psi^{-}\rangle = \frac{1}{\sqrt{2}}\left(|\uparrow_n\rangle_A |\downarrow_n\rangle_B - |\downarrow_n\rangle_A |\uparrow_n\rangle_B\right),
\end{equation}
where the kets $\left\{|\uparrow_n\rangle, |\downarrow_n\rangle\right\}$ are the eigenstates of a single-qubit Pauli operator $\sigma_n\coloneqq\bm{\sigma}\cdot \bm{n}$ locally defined at each one of the parties' labs ($A$ and $B$). The expectation value of any single-qubit observable with respect to the singlet state is zero,
\begin{align}
    \bra{\psi^{-}} \sigma_m\otimes\mathds{1} \ket{\psi^{-}}= \bra{\psi^{-}} \mathds{1}\otimes\sigma_{m'} \ket{\psi^{-}}=0,
\end{align}
while the correlation between any pair of observables measured each on one of the qubits $\sigma_m\otimes\sigma_{m'}$ is determined by their relative orientation as, 
\begin{align}\label{eq:cos}
\bra{\psi^{-}} \sigma_m\otimes\sigma_{m'} \ket{\psi^{-}} =-\bm{m}\cdot\bm{m}'=-\cos(\theta).
\end{align}
This equation implies that the orientation of any one of the detectors may always be taken as a reference for defining the orientation of the other one (or, more precisely, with respect to their orientation at which their outcomes are fully anti-correlated). 

One key feature that we wish to emphasize here is that all two-qubit maximally entangled states are related to each other by local operations. This observation, which is a direct manifestation of the LOCC theorem for entanglement transformation~\cite{Nielsen99}, can be readily shown by mapping a general maximally entangled state 
\begin{align}
    \ket{\Psi}=\frac1{\sqrt{2}}(\ket{\uparrow_n}_A\ket{\uparrow_{n'}}_B+e^{i\Phi}\ket{\downarrow_n}_A\ket{\downarrow_{n'}}_B),
\end{align}
to $\ket{\psi^{-}}$ using a local unitary transformation (for example, on qubit $B$) $U(\bm{n},\bm{n}',\Phi)$ that takes $\{\ket{\uparrow_{n'}}, e^{i\Phi}\ket{\downarrow_{n'}}\}$ to $\{\ket{\downarrow_n}, -\ket{\uparrow_n}\}$. This leads to the conclusion that given a physical source of two-qubit maximally entangled state, there is always a joint reference frame (in the Hilbert space of the two qubits) in which we can describe the state as a maximally entangled state of our choice -- particularly as the singlet state. 

The implication of this fundamental result is that we can rewrite the correlation, Eq.~\eqref{eq:cos}, as: 
\begin{align}
 \bra{\psi} \sigma_{m}\otimes\sigma_{m''} \ket{\psi} = -\cos(\theta),
\end{align}
where $\bm{m}\cdot\bm{m}''=\cos(\vartheta)$, for some $\vartheta\in[-\pi,\pi)$, and $\ket\psi$ is a maximally entangled state of the form
\begin{align}
    \ket{\psi}=\frac1{\sqrt{2}}(\ket{\uparrow_n}\ket{\downarrow_n}+e^{i\Phi}\ket{\downarrow_n}\ket{\uparrow_n}),
\end{align}
as long as $\theta=\Phi+\vartheta$ and the orientations $\bm{m}$, $\bm{m}'$ and $\bm{m}''$ are perpendicular to $\bm{ n}$.

\section{Extending the model to events distributed on the two-sphere}\label{app:extSph}
In this section we  discuss in broad strokes a possible extension of our statistical model to the case where the events observed by Alice's and Bob's detectors are location on the two-sphere rather than the unit circle. This extension is desirable since the quantum mechanical treatment of spin-$1/2$ particles (or of the polarization degree of freedom of photons)  is given by elements from the special unitary group $SU(2)$, which is a double cover of the three-dimensional rotation group $SO(3)$ (a complete consideration of this extension will be presented in a future publication). 

In this extension, similar to the model we introduced in the main text, in the absence of an absolute frame of reference, one cannot assign distinct physical degrees of freedom to the orientation of the emitted particles without relating them to the orientation of the corresponding detectors. Thus we label the emitted particles only through their relative orientation to each one of the detectors, $A$ and $B$, with the help of corresponding random variables defined over the two-sphere. 

In the model discussed in the main text, the probability density of events over the unit circle $\rho(\Delta)$ was not necessarily uniform (though it had to have specific symmetry properties), and the transformation rule $\Gamma(\Delta)$ defined by Eq.\eqref{eq:trinity} was a density-preserving map over the unit circle. This ensured that the probability density of events is preserved when transforming from one observer to the other. In the generalized model, we follow the same concept. However, due to the symmetry consideration on the two-sphere, we find interestingly, that it suffices to consider uniform distributions of events. In this case we similarly require that any mapping of events between the two observers must preserve the uniform distribution over the two-sphere.  Therefore in this generalized setting, the relevant operation is the group of area preserving diffeomorphisms (of the double cover) of the two-sphere. This group is a family of (continuous, differentiable) transformations that map uniform distributions to uniform distributions, that include, specifically, transformations that produce the quantum correlations $E(\theta)=-\cos \theta$.  As mentioned above, a full consideration of this model is under development, and will be given elsewhere.
\end{document}